\documentstyle[twocolumn,aps,prl,epsf]{revtex}

\begin{document}

\title{SCALING OF FOLDING PROPERTIES IN SIMPLE MODELS
OF PROTEINS}

\author{Marek Cieplak, Trinh Xuan Hoang and Mai Suan Li}

\address{Institute of Physics, Polish Academy of Sciences,
Al. Lotnikow 32/46, 02-668 Warsaw, Poland }
\address{
\centering{
\medskip\em
{}~\\
\begin{minipage}{14cm}
Scaling of folding properties of proteins is studied in a toy system -- the
lattice Go model with various two- and three- dimensional geometries of the
maximally compact native states.  Characteristic folding times grow as power
laws with the system size. The corresponding exponents are not universal.
Scaling of the thermodynamic stability also indicates size-related
deterioration of the folding properties.
{}~\\
{}~
\end{minipage}
}}

\maketitle
\newpage

Recent advances in understanding of protein folding have been made, to
a large extent, through studies of lattice heteropolymers with a small
number of beads, $N$, \cite{Dill,Sali}.   In these toy models of
proteins, the beads represent aminoacids.  Lattice models allow for an
exact determination of the native state, i.e. of the ground state of
the system, and are endowed with a simplified dynamics.   An $N$ of
order 125 is considered to be large in such studies and then special
sequences are considered \cite{Dinner}.  There are real life proteins
\cite{Creighton} with $N$ as small as of order 30, but most of them
are built of several hundreds aminoacids. Apparently there is no
protein with $N$ exceeding 5000 which is orders of magnitude smaller
than the number of base pairs in a DNA.  The question we ask in this
Letter is: how do folding properties of proteins scale with $N$ and
can they lead to a deterioration in stability and kinetic
accessibility of the native state that exceed bounds of functionality?

A previous numerical analysis of the scaling has been done by Gutin,
Abkevich, and Shakhnovich \cite{Shakhnovich} who studied three
dimensional (3$D$) lattice sequences with $N$ up to 175.  For each
$N$, they considered 5 sequences and selected one that folded the
fastest under its optimal temperature $T_{min}$.  The corresponding
folding time, $t_{01}$, was the quantity that was used in studies of
scaling.  They discovered that $t_{01}$ grows as a power law with the
system size:
\begin{equation}
t_{01} \;\sim\; N^{\lambda} \;\;.
\end{equation}
The exponent $\lambda$ was found to be non-universal -- it depended
on the kind of distribution of the contact energies $B_{ij}$ in the
Hamiltonian
\begin{equation}
H\;=\;\sum _{i<j} \; B_{ij}\; \Delta _{ij} \;\;\; ,
\end{equation}
which pointed to existence of a variety of kinds of the energy
landscapes \cite{Bryngelson}. In eq.(2), $\Delta_{ij}$ is either 1 or
0 depending on whether the monomers $i$ and $j$ face each other, but
not along the chain, or not.  For  random and designed sequences, with
the $B_{ij}$'s generated from the data base of Ref. \cite{Miyazawa},
$\lambda \approx 6$ and $\approx 4$, respectively \cite{Shakhnovich}.
Finally, for the Go model \cite{Go}, in which $B_{ij}=-1$ for native
contacts and 0 for non-native contacts, $\lambda \approx 2.7$. 
There were also
phenomenological arguments \cite{Wolynes} which suggested that the
folding times scale with $N$ exponentially for all temperatures. Thus
the nature of the scaling laws for the folding times remains puzzling.
Perhaps more importantly, Gutin et al. \cite{Shakhnovich} did not study
scaling of any of the characteristic temperatures that are relevant
for folding nor the effects of the dimensionality were explored.

In this Letter, we report on studies of the 2 and 3$D$ Go model, with $N$ up
to 56 and 100 respectively. In the 2$D$ and $N$=16 case,  
we consider all 37 maximally compact conformations 
(there are 69 such conformations
but only 38 of them are distinct due
to the end-to-end symmetry of the Go model; furthermore, one 
conformation cannot be accessed kinetically). In the remaining cases,
we study 15 conformations, except for $N$=80 and 100 
when only 10 and 5, respectively, are considered.
Note that each of these structures is equally designable within the model
because each is a nondegenerate ground state to one Go sequence. 
We demonstrate that in this case, $t_{01}$ is indeed given
by eq. (1). In 2$D$,
$\lambda$ is $5.9 \pm 0.2$ 
Thus the constraint for
the heteropolymer to lie in a plane increases $\lambda$ compared to
the 3$D$ Go model.
Our larger statistics also allows us to study median
values, not just minimal, of the folding times. The median values also
follow the power law with an effective $\lambda$ of $6.3 \pm 0.2$
and $3.1 \pm 0.1$ in 2 and 3$D$ respectively.
Actually, the effective $\lambda$ depends on whether the folding is
studied at $T_{min}$ or at the folding temperature $T_f$.   $T_f$ is
defined operationally as a temperature at which the equilibrium
probability of finding the native state is $\frac {1}{2}$.   We find
that in 2$D$ and at $T_f$, $\lambda \;=\;6.6 \pm 0.1$  (the exponent for the
minimal time at $T_f$ is $6.3 \pm 0.3$) which means that by moving
away from conditions which are optimal for the folding kinetics
one generates a somewhat increased exponent in the power law.
 
Notice that good folding takes place for sequences for which $T_f$ is
comparable to or bigger than $T_{min}$.  Otherwise the folding is
poor. An important novel aspect of of our research is that we
determine the scaling properties of $T_{min}$ and those of the folding
temperature, $T_f$.  We conclude that, both in $2$ and $3D$, there are
indications that there could be a size related limit to good
foldicity.   We find that $T_{min}$ grows linearly with $N$ whereas
$T_f$ first grows like $T_{min}$ but then it falls off and possibly
saturates asymptotically. This makes the gap between $T_{min}$ and
$T_{f}$ increase linearly with $N$ asymptotically which would change
the folding kinetics from excellent to bad.

One stumbling block in studies of scaling of random systems is the
necessity to compare quantities which are averaged statistically and
to have some control of the statistical ensemble used.  The advantage
of the Go model is that there is no randomness in the values of the
contact energies and the ensemble is generated by the set of possible
maximally compact conformations that can act as native states -- i.e.
the variety is only due to the geometry of the native states.  The
advantage of studying 2$D$ models is that, for $N$=16, it is feasible
to determine the full distribution of $T_f$, $T_{min}$, and of the
folding times among all of the 37 targets and then to realize that
the {\it median} folding time probes vicinities of the peak in the
distribution. Thus, on going to larger $N$ and taking, as we usually do,
15 targets, it is reasonable to expect that the corresponding median time
still probes the peaks of good foldicity.  Median quantities are, in
addition, more stable statistically, in general, whenever one deals
with wide distributions.

\begin{figure}
\epsfxsize=3.2in
\centerline{\epsffile{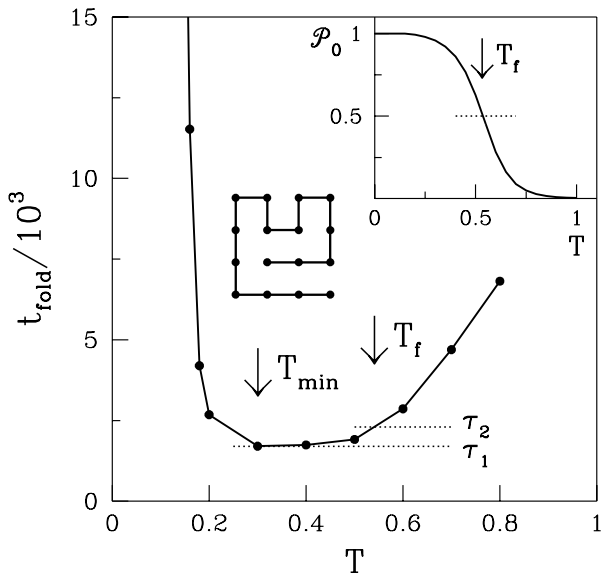}}
\caption{The dependence of the median folding time, $t_{fold}$,
on $T$ for the 2$D$ Go conformation shown in the center of the figure.
The results are averaged over 1000 Monte Carlo trajectories.
The inset shows the $T$-dependence of probability for the sequence to be
in the native state as obtained through an exact evaluation of
the partition function which involves 802075 conformations.}
\end{figure}

As to the selection of the 15 native maximally compact 
targets: in 2$D$ 10 were obtained by a
random construction and 5 were obtained by a multiple quenching of
randomly shaped homopolymers until a maximally compact conformation
was obtained. The homopolymers had identical attraction in each
possible contact. In both methods, we generate targets to which there
is a path of kinetic access. In 3$D$, all targets were obtained by
the random construction.

Figure 1 illustrates definitions of quantities that will be studied
here.  It shows the dependence of the median folding time, $t_{fold}$,
on temperature for one target.  The target has $N$ of 16 and is shown
in the center of the figure.  The optimal temperature, $T_{min}$, is
where $t_{fold}$ is the shortest.  $T_{min}$ signifies the onset of
glassy kinetics. This quantity is better suited to study scaling than
the glass transition temperature $T_g$ \cite{Socci} because the latter
involves a cutoff time which necessarily must be $N$ dependent.
$t_{fold}$ at $T_{min}$ will be denoted by $\tau _1$.   $\tau _2$ is
defined to be $t_{fold}$ at $T_f$ ($T_f$ is larger than $T_{min}$ for
the target shown in Figure 1).   In the statistical ensemble, $t_1$ is
defined to be the median value of $\tau _1$ and $t_2$ -- the  median
value of $\tau _2$. We also study $t_{01}$ and $t_{02}$ which are the
minimal values of $\tau _1$ and $\tau _2$ among the targets
considered.

\begin{figure}
\epsfxsize=3.2in
\centerline{\epsffile{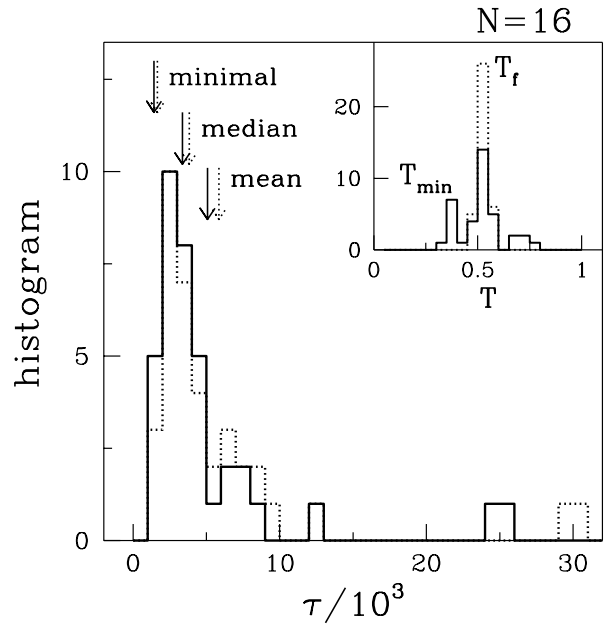}}
\caption{The distribution of folding times at $T_{min}$
(solid lines) and at $T_f$ (dotted lines) for the 2$D$ $N$=16 case.
The arrows indicate their median, mean,  and minimal values.
The inset shows the distributions of $T_{min}$ and $T_f$.}
\end{figure}

The folding times were obtained through a Monte Carlo procedure that
satisfies the detailed balance condition \cite{Malte}, and was
motivated by studies presented in Ref. \cite{Chan}.   For each
conformation of the polymer, one first determines the number of
possible single and double-monomer (crankshaft) moves -- these numbers
will be denoted here by $A_1$ and $A_2$ respectively.  The maximum
value of $A_1 + A_2$, among all conformations, is equal to
$A_{max}=N+2$. Probability to attempt a single monomer move is taken
to be $rA_1/A_{max}$ ($r$=0.2).  For a double monomer move it is
$(1-r)A_2/A_{max}$. The attempts are rejected or accepted as in the
standard Metropolis method.  The folding time is defined as 
the first passage time and is measured by the 
number of Monte Carlo attempts divided by $A_{max}$. For $N>16$, it is
determined based on 50 to 200  trajectories.
It should be noted that ref.\cite{Shakhnovich} does not specify
whether the detailed balance condition was enforced.

\begin{figure}
\epsfxsize=3.2in
\centerline{\epsffile{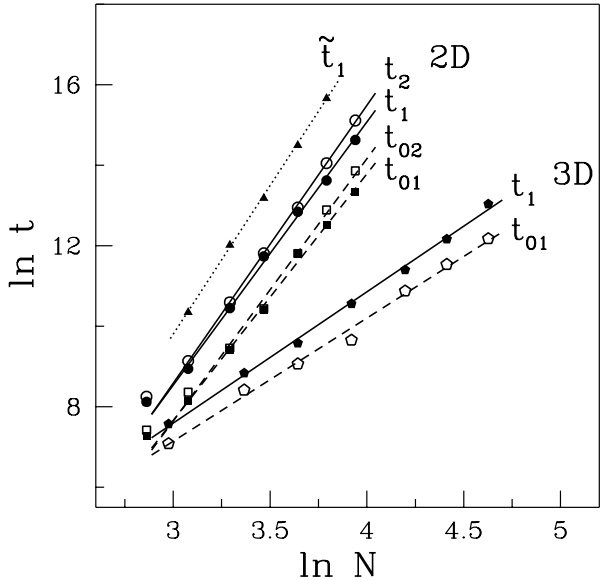}}
\caption{The dependence of folding times on $N$. $t_1$ and $t_2$ are
the median folding times at $T_{min}$ and $T_f$ respectively.
$t_{01}$ and $t_{02}$ are the corresponding minimal folding times found.
$\tilde{t}_1$ is the median folding time at $T_{min}$ as obtained
by a straightforward Monte Carlo procedure 
which does not enforce the detailed balance condition.
The values of the effective exponent $\lambda$ for the 2$D$ case are:
$7.1 \pm 0.1$, $6.6 \pm 0.1$, $6.3 \pm 0.2$, $6.3 \pm 0.3$, and
$5.9 \pm 0.2$ when counting clockwise. For 3$D$, the slopes are
$3.1 \pm 0.1$ and $2.9 \pm 0.1$.
}
\end{figure}

Figure 2 shows the distribution of $\tau _1$ and $\tau _2$ for all
targets with $N$=16. There is a substantial scatter in the values of
$\tau _i$ so the usage of the median $t_i$ appears to be justified.
The inset shows the corresponding distributions of $T_f$ and
$T_{min}$.  Both are centered and the median and mean values almost
coincide.  Note that there is very little variation in $T_f$: all Go
targets with $N$=16 have almost identical stability properties: $T_f$
varies between 0.489 and 0.565. On going to larger $N$'s, the
distributions of $\tau _1$ remain clustered around $t_1$ but the long
time tail appears to extend towards longer and longer times. This
results in an overall flattening  of the distributions on the scale
set by $t_1$.  For $N$=16, the exact distribution of $\tau _1/t_1$
ends at about 8 whereas  our sampling of $N$=20 and 42 yields tails in
$\tau _1/t_1$ which are located at around 16 and 10 respectively.
Within our statistics, we have not spotted any relatively long lasting
folding processes for other values of $N$.  However, their very
existence for $N$=20 and 42 suggests an emergence of the tails in
distributions if those could be sampled fully.

Figure 3 summarizes our results on the scaling of folding times.  It
demonstrates the validity of the power laws both for the median and
for the minimal folding times. The effective exponents $\lambda$
depend on $T$, i.e., they depend on whether the kinetics was monitored
at $T_{min}$ or $T_f$. This dependence is not substantial but it
indicates variations of the free energy landscape with $T$ and
underscores a more general lack of universality.

The generic power laws obtained by Gutin {\em et. al.}\cite{Shakhnovich} and
by us contradict the
exponential laws derived in the random energy model
\cite{Bryngelson,Wolynes}. They support a generally accepted view that
the folding process is a finite volume version of the first order
transition \cite{Thirumalai,Shakhnovich}.  In this picture one may
visualise the transition stage as an inhomogeneous mixture of the
"new" phase in the sea of an "old" phase \cite{Lifshitz}.  The random
energy model does not capture such inhomogeneities.

\begin{figure}
\epsfxsize=3.2in
\centerline{\epsffile{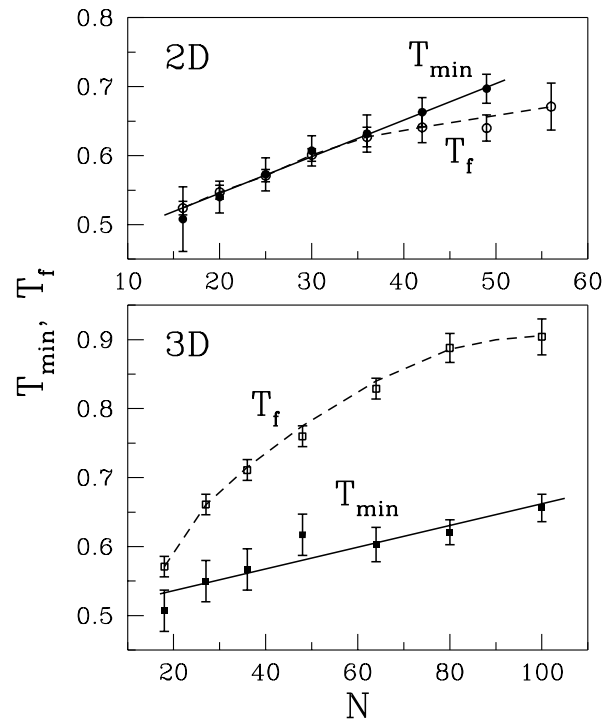}}
\caption{The dependence of $<T_{min}>$ and $<T_f>$ on
$N$. $<T_{min}>$ is fitted by a linear function:
$\;0.44 + 0.0053 N$ and $ 0.505 + 0.0015(7) N$ for 2 and 3$D$ respectively.
The results are averaged over the conformations that were used
in the studies of dynamics.}
\end{figure}

Figure 4 shows the $N$-dependence of the characteristic temperatures.
For both 2 and 3 $D$ $T_{min}$ grows linearly with $N$ whereas
$T_f$ shows a more complex behavior.
For $N > 16$, $T_f$ is determined from the Monte Carlo simulations: 
a) we vary the $T$ in steps of 0.05 or smaller, b) at each $T$ we start
from the native state and monitor the probability of occupying it,
c) in most cases the results are averaged over 50 different trajectories.
The number of Monte Carlo steps for each $T$
depends on $N$ and it ranges from $10^6$ to 7$\times 10^6$. We checked
that doubling the selected cutoff times had negligible effect on
$T_f$. The procedure yields results which agree with those obtained by
the exact enumeration for $N$=16. In 2$D$, the dependence of $<T_f>$ on $N$
initially follows that of $T_{min}$. However, on crossing $N_c$ of 36,
$T_f$ falls off and it may saturate which is suggested by the
declining rate of growth.  Thus 2$D$ Go conformations appear to have
intrinsic limits to their thermodynamic stability. Beyond $N_c$, the
foldicity becomes gradually poorer and poorer. The same scenario
appears to be present also in the 3$D$ case except that the small $N$
value of $T_f$ is substantially larger than $T_{min}$.
$T_f$ starts showing signs of the saturation around $N$=80.
We were unable to explore values of $N$ that were larger than 100 but
a saturation of $T_f$ is expected on general grounds due to the existence
of the (first order) phase transition to the folded phase
in the thermodynamic limit. $T_{min}$, on the other hand,
is expected to grow indefinitely due to the growth of kinetic barriers
to cross.  In 3$D$, $T_f$ and $T_{min}$ appear to cross somewhere around
$N_c$=300.

In conclusion, we have studied the scaling properties not only of the
fastest sequences, as in ref.\cite{Shakhnovich},
but also of those with typical folding rates.
The exponents in the resulting power laws for the  folding times
depend on $D$, values of the $B_{ij}$'s, and on $T$. In addition to
the deterioration of the folding kinetics with $N$, as described by
the growth of $T_{min}$ and of the folding times, a relative
deterioration of the thermodynamic stability also appears to set in.
Thus there will be no rapidly folding heteropolymers of a large size.
It would be interesting to determine the scaling properties for more
realistic models of proteins.

This work was supported by KBN (Grant No. 2P03B-025-13). Fruitful
discussions with Jayanth R. Banavar and Dorota Cichocka are gratefully
acknowledged.


\vspace{0.5cm}

\begin{references}

\bibitem{Dill} K. A. Dill {\em et al}, Protein Science {\bf 4}, 561
(1995).

\bibitem{Sali} A. Sali, E. Shakhnovich, and M. Karplus,J. Mol. Biol.
{\bf 235}, 1614-1636 (1994)

\bibitem{Dinner} A. R. Dinner, A. Sali, and M. Karplus, 
Proc. Natl. Acad. Sci. USA, {\bf 93}, 8356-8361 (1996).

\bibitem{Creighton} See e.g. T. E. Creighton, 
{\it Proteins: Structures and Molecular Properties},
W. H. Freeman and Company, New York, 1993.

\bibitem{Shakhnovich} A. M. Gutin, V. I. Abkevich, and
E. I. Shakhnovich, Phys. Rev. Lett. {\bf 77}, 5433 (1996)

\bibitem{Bryngelson} J. D. Bryngelson, J. N. Onuchic, N. D. Socci, and
P. G. Wolynes, Proteins {\bf 21}, 167 (1995).

\bibitem{Miyazawa} S. Miyazawa and R. l. Jernigan, Macromolecules 
{\bf 18}, 534 (1985).

\bibitem{Go} N. Go and H. Abe, Biopolymers {\bf 20}, 1013 (1981).

\bibitem{Malte} M. Cieplak, M. Henkel, J. Karbowski, and J. R. Banavar,
Phys. Rev. Lett. {\bf 80}, 3654 (1998);
M. Cieplak, M. Henkel, and J. R. Banavar, J. Cond. Matt. (in press).

\bibitem{Wolynes} J. D. Bryngelson, and
P. G. Wolynes, J. Chem. Phys. {\bf 93}, 6902 (1989); E. Shakhnovich
and A. M. Gutin, Europhys. Lett. {\bf 9}, 569 (1989); J. Saven,
J. Wang, and P. Wolynes, J. Chem. Phys. {\bf 101}, 11037 (1994).

\bibitem{Socci} N. D. Socci and J. N. Onuchic, J. Chem. Phys.
{\bf 101}, 1519 (1994); see also M. Cieplak and J. R. Banavar,
Fold. Des. {\bf 2}, 235 (1997).

\bibitem{Chan} H. S. Chan and K. A. Dill, J. Chem. Phys.
{\bf 99}, 2116 (1994); H. S. Chan and K. A. Dill, J. Chem. Phys.
{\bf 100}, 9238 (1994).

\bibitem{Thirumalai} D. Thirumalai, J. Phys. I (France) {\bf 5}, 1457 (1995).

\bibitem{Lifshitz} E. M. Lifshitz and L. P. Pitaevskii, {\em Physical
Kinetics} (Pergamon, London, 1981).

\end{references}
\end{document}